\documentclass[12pt]{article}
\input epsf.tex

\usepackage{latexsym}
\usepackage{graphicx}
\usepackage{amsmath,amstext}
\usepackage{comment}
\numberwithin{equation}{section}
\newcommand{\be}{\begin{equation}}
\newcommand{\ee}{\end{equation}}
\newcommand{\benn}{\begin{equation*}}
\newcommand{\eenn}{\end{equation*}}
\newcommand{\bea}{\begin{eqnarray}}
\newcommand{\eea}{\end{eqnarray}}
\newcommand{\bean}{\begin{eqnarray*}}
\newcommand{\eean}{\end{eqnarray*}}

\begin{document}
\begin{titlepage}

\vbox{
      \hbox{SCIPP-08/02}
      \hbox{April, 2008}
}
\vspace*{2cm}

\begin{center}

\vskip0.2in

{\Large \bf New Light Windows for Sparticle Masses and Higgs Decays in the R Parity Violating MSSM $\footnote{This paper is dedicated to the memory of E.J. Rhee}$}

\vskip 0.3 in

Linda M. Carpenter$^{\dagger}$ David E. Kaplan$^{\dagger \dagger}$ E.J.Rhee$^{\dagger \dagger}$

\vskip0.2in

\emph{$^{\dagger}$ UC Santa Cruz\\Santa Cruz CA}

\emph{$^{\dagger \dagger}$ Johns Hopkins University\\Baltimore MD}

\begin{abstract}


In supersymmetric models with R parity violation, constraints on superpartner masses are significantly weaker than in models which conserve R parity.  We find in regions of parameter space where a neutral gaugino or third generation scalar is decoupled from the Z and/or traditional gaugino mass relationships do not hold allow some particles to be light enough to allow them to be the decay products of the Higgs.  For example, surprisingly a stau could be lighter than 30 GeV if it decays hadronically.  We estimate the Higgs bounds when one of these decays dominates and find allowed Higgs masses well below the current LEP bound on the standard model Higgs.  We also survey the rich variety of final states in Higgs production.

\end{abstract}
\end{center}
\end{titlepage}

\section{Introduction}


Softly broken supersymmetry (SUSY) is arguably our best candidate for an explanation of what cuts off the quadratic divergent quantum corrections to the squared Higgs mass in the standard model (SM).  Because of these corrections, the SM alone, with a cutoff much higher than the weak scale, suffers from a fine-tuning problem.  The mass scale of superpartners effectively represent the cutoff and above that scale the tuning is no longer necessary.  Thus, to completely remove tuning from the Higgs, the natural scale for superpartner masses is on the order of the weak scale.



Of course neither the Higgs superpartners have been discovered.  The bound on the SM Higgs mass is 114.4 GeV \cite{LEPhiggs}, and this bound, to a good approximation, applies to the lightest Higgs in most of the parameter space of the so-called 'mSUGRA' version of the minimal supersymmetric standard model (MSSM) \cite{lepSUSYhiggs}.  These bounds are strong enough to require fine tuning within the mSUGRA model, the most restrictive being the Higgs mass bound - the physical Higgs mass in mSUGRA lies naturally below the $Z$ mass, and increases above the $Z$ with contributions only logarithmically sensitive to the superpartner mass scale.  Superpartner masses (mainly the stop mass) need to be pushed beyond the electroweak scale in order to satisfy the bound.


Going beyond mSUGRA, Higgs and superpartner mass bounds can change - and sometimes radically.  Here we study a version of the MSSM with R-parity violation.  We also consider regions where the standard gaugino mass relations do not hold.  The main phenomenological feature of violating R-parity is that the lightest superpartner is no longer stable and that 'missing energy' is either strongly reduced or eliminated in events with superpartners produced.  This single change, allowing the lightest supersymmetric particle (LSP) to decay, reduces bounds on nearly every superpartner (save the chargino) to below 100 GeV, and sometimes well below.  This keeps open the possibility of the Higgs boson decaying into superpartners.


In this paper, we explore Higgs decays into a pair of LSPs which result in multi-body final states.  If the Higgs decays in a non-SM way, it affects the lower bound on the Higgs mass as the standard LEP searches are less efficient or simply do not apply.  Based on a wide range of searches, we estimate that many of these new decays reduce the lower bound on the Higgs mass to 105 GeV or less.  While this seems like a small change, the exponential sensitivity of the superpartner ({\it e.g.}, stop) masses to the needed Higgs mass correction makes this reduction relevant for re-opening SUSY parameter space. The key is beating the standard decays of $h \rightarrow b\overline{b}$ by a large amount - and this happens in broad regions of parameter space the Higgs sector is in the decoupling limit.

This paper is organized as follows, section 2 reviews R parity violating operators and their bounds. Section 3 discusses constraints, signatures and parameter space for neutral gaugino LSPs.  Section 4 discusses constraints, signatures and parameter space for third generation scalar LSPs, and section 5 concludes.  In the appendix, we discuss the fit to $e^+ e^-$ hadronic cross section data when a new light charged scalar is in the spectrum.

\section{R parity violation}

In the MSSM generally one imposes a symmetry known as R parity whereby fields and superfields are given opposite parity.  Only operators with positive parity are allowed to appear.  The LSP is stable; if one starts with N superparticles, one must end with N mod 2 due to their negative parity. However we are free to introduce parity violating operators into the superpotential:
\begin{eqnarray}
 W & \supset &
 \mu_i L_i {\bar H} + \lambda_{ijk} L_i L_j E^c_k + \lambda'_{ijk} L_i Q_j D^c_k \nonumber \\ & + & \lambda''_{ijk} U^c_i D^c_j D^c_k ,
 \label{eq:rpv}
 \end{eqnarray}
where $L$, $E^c$, $\bar{H}$, $Q$, $U^c$, and $D^c$ are lepton doublet, lepton singlet, up-type Higgs, quark doublet, up-type quark singlet, and down-type quark singlet superfields respectively, and the $ijk$ are flavor indices.  The first operators violate lepton number conservation, while the third violates baryon number conservation.  An acceptable theory cannot have both LNV and BNV couplings of non-negligible size as it would cause rapid proton decay.

For a general flavor of RPV bounds see \cite{Allanach:1999ic,Bhattacharyya:1997vv}.  In general LLE operators are of order a few times $10^{-2}$, while LQD and UDD operators are bounded at $10^{-1}$ though a few couplings have much harsher bounds and some are order 1.  LNV bounds come from a great variety of sources.  Semi-leptonic meson decays put bounds on a many products of couplings \cite{Dreiner:2006gu} most of which fall in the range $10^{-2}-10^{-4}$.  The most stringent bound for LLE operators is on the $\lambda_{133}$ operator and comes from contributions to the electron neutrino' majorana mass and is less than $10^{-3}$ for 100 GeV superpartner masses.  Other LLE bounds come from measurements of $R_{\tau}$ and $R_{\tau \mu}$.
The LQD constraints come from a variety of sources.  Neutrinoless double beta decay
puts a strict bound on the $\lambda_{111}^{'}$ operator of $\sim 10^{-4} \times (m_{\chi}/100GeV)^{1/2}(m_{\tilde{e}}/100GeV)^2$.  Charged current universality, $D_s$ and $D$ meson decays and the measurement of $R_{\tau}$ also place bounds and in general operators involving third generation couplings $\lambda_{331}^{'}, \lambda_{332}^{'},\lambda_{333}^{'}$ and $\lambda_{232}^{'}$ are relatively unconstrained.
The strictest BNV bounds come from  double nucleon decay and apply to first generation couplings $\lambda_{112}^{''} \sim 10^{-7}$.  Neutron anti neutron oscillation limits $\lambda_{113}^{''}$ to $\sim 10^{-4}$.  Other couplings are less constrained and bounds vary from $10^{-1}$ to 1.  A generally safe statement is that with the exception of a few pathological couplings cited in the canon, nearly all bounds allow for reasonably prompt decay of a superpartner into SM particles.  Avoiding stringent constraints on some of the first generation RPV operators may be remedied easily by requiring a Yukawa-hierarchical coupling scheme where third generation couplings are dominant; this scheme will be of particular value when considering LNV decays of gaugino LSPs.  Here we favor a Yukawa hierarchical scheme and consider one only one RPV coupling at a time to be turned on.  We now go on to list possible LSPs which both may be lighter than half the Higgs mass, and then may decay via RPV.

\section{Gaugino LSPs}
\subsection{Topology and General Constraints}
The Higgs may decay into a pair of neutral gaugino LSPs.  The RPV decay then proceeds as each gaugino decays to a fermion and virtual sfermion which itself decays via an RPV vertex to two SM particles.  It would be simple here to think of gauginos decaying through an effective four fermi interaction to three fermions $\chi_0 g_i\lambda_{jk}/M_{\overline{f}}$.  The topology of these decays is Higgs to 6 SM fermions.  In the case of BNV we would have Higgs to six jets.  For LNV, Higgs to 4 jets plus 2 leptons, and LLE to 6 leptons.

There are several general constraints on this scenario applicable to all gaugino LSPs.
The first is the mass requirement that gauginos be less than half of the Higgs mass.  What usually stands in the way of gauginos this light are assumptions about mass relations in standard SUSY mass schemes.  Gauge mediation, anomaly mediation, and MSUGRA relate the gauginos by a single mass parameter.  There is a tight experimental lower bound on the chargino mass of 102.7 GeV, which then constrains the mass of the lightest neutralino through standard gaugino mass relations.  In addition, most standard mass schemes also predict a gluino much heavier than the other gauginos.  For example, MSUGRA predicts the M2:M3 ratio to be about 3:1 thus a chargino passing the lower mass bound implies a heavy gluino.  By rejecting the standard gaugino mass relationships we open up large regions of SUSY parameter space.

We must also address lower mass bounds on sparticles from existing RPV searches.  DELPHI and L3 have put lower bounds on the neutralino as the LSP decaying through RPV \cite{Abdallah:2003xc} \cite{Achard:2001ek} \cite{Acciarri:2000nf}.  L3's mass limits for $\chi_0$ decaying via LQD is 32.6GeV, while it's limits on decay via LLE and UDD are 40.2GeV and 39.9GeV respectively.  DELPHI has put lower mass limits on decay via LLE and UDD as 39.5GeV and 38GeV.  At DELPHI, a simple counting experiment was done assuming gaugino mass unification. We will examine the LLE decay count, since it was the lowest; if we avoid the LLE constraint we avoid them all.  In the LLE channel DELPHI found only 1.5 events above the background at 95 percent confidence.  Recalling that $N= \epsilon L \sigma $ where $\epsilon$ is the efficiency, we see that for $\epsilon$ between .11 and .38 and an integrated luminosity of 437.8$pb^{-1}$, the production limit is $\sim$ .03-.01 pb.  If we assume no contribution from charginos, a selectron mass of between 380 and 500 GeV will sufficiently suppress this process for a 30 GeV neutralino.  There were additional searches at ALEPH that placed lower bounds of 29 GeV and 23 GeV on LQD and LLE decays of neutralinos respectively \cite{Barate:1997ra}
\cite{Barate:1998gy}.  These analyses also relied on MSUGRA.  For example, they put an upper bound on neutralino production from LLE decay which is at least .5pb.  Without MSUGRA, we may easily beat this production bound by picking the selectron mass.

In the case of gluino as the LSP no direct decay bounds are quoted.  An indirect search exists  which looks for direct decay of a neutralino produced in the decay of a squark, and assuming gaugino mass unification \cite{Abazov:2001nt}.  This search concerns the coupling $\lambda_{2jk}^{'}$ only and looks for muons in the final sate. For reason which will be explained this does not fall into our scenarios.  Interestingly, a lower mass limit on the gluino is set at 6.3 GeV \cite{Janot:2003cr}. This is obtained from the contribution of hadronic decay width of the Z when $e^+ e^- \rightarrow q\overline{q}$ and a quark radiates a gluon which decays to a gluino pair.

The decay of Higgs to neutralinos must also beat the standard Higgs decays if current searches are not to have already ruled out the Higgs.  The largest standard decay rate is Higgs$\rightarrow b\overline{b}$.  The bottom Yukawa coupling is not particularly large and we shall show that it does not require special fine tuning in parameter space in order for Higgs decays to gauginos to beat the bottom decay by a factor of 5 or so.

The next constraints heavily involve the couplings to Z bosons.  Decays of light neutral gauginos may not substantially  change the Z width, nor may they have a large effect on the predictions for total e+ e- to hadrons cross section.  These issues may be solved by decoupling the lightest neutralino sufficiently from the Z, and making sure loop induced couplings from of the Z to the gluino are not too great.

There is a constraint from supersymmetric contributions to $b \rightarrow s\gamma$ which come mainly from the charged Higgs and the chargino (diagrams involving the gluino also make contributions but are suppressed by insertions of the CKM matrix).  Diagrams with the $ \chi ^{+}$ and $H^+$ have a stop in the loop.   Avoiding large contributions to this process  is accomplished in two ways: if the stop and the charged Higgs are heavy and all diagrams are suppressed, or if the diagrams cancel.  Often there is some tension between a heavy stop, which could tune the Higgs sector, and small $b- s \gamma$.  The current experimental bound is $3.55 +- .24^{+.09}_{-.1} +- .03 x 10^{-4}$\cite{Barberio:2007cr}.  The current Standard Model theoretical prediction made at NNLO is now significantly below the measurement at $3.15 +- .23 x 10^{-4}$\cite{Misiak:2006zs}, so there is some small space for supersymmetric contributions.

There are generational constraints on LNV decays which have charged leptons in the final state.  Tight constraints from Tevatron exist on like sign dilepton signals \cite{Abulencia:2007rd}.  This search found a slight excess of like sign dilepton events, 20 events at 95 percent confidence, at an integrated luminosity of 1$fb^{-1}$.  The efficiencies for seeing new physics that produced a WZ were assumed to be around 8 percent.  In SUSY optimized scenarios the observed excess was only 8 events and the expected efficiency was slightly lower but this scenario required a large transverse momentum imbalance which presents difficulties for a symmetric decay.  If we consider that the standard Higgs production cross section at Tevatron is over $10^3$ fb, we see that if the generational LLE couplings were the same we would have expected at least a 40 event excess in like sign dilepton events; as two neutralinos decaying through a virtual charged slepton will produce like sign dileptons half the time.  Taus were not covered in the dilepton search since they looked for well isolated lepton pairs that could be tracked back to a single vertex.   This tells us that the LNV operator must favor third generation processes in order to avoid constraints.

Finally, since these decays involve virtual sfermions and small couplings they may not be prompt.  The general decay length has the form
\be
L \sim \frac{Nm_{\tilde{m}}^4}{c^2 \lambda^2 m_{\chi}^5}  \beta \gamma
\ee
where $\lambda$ is the RPV coupling, $m_{\tilde{m}}$ is the sfermion mass and N is a factor specific to the decay.  If the decay length is long enough, the decay will have two displaced vertices that are separated from the primary decay vertex.  However if the decay length is too long, the decay will occur outside of the detector and will thus be ruled out by existing missing energy searches.  In this way we may put bounds on the RPV coupling and sfermion masses.  We go on to do specific analyses of gaugino decays.

\subsection{Neutralino LSP}

The possibility of a light neutralino LSP decaying via BNV was discussed in detail in ref \cite{Carpenter:2006hs}.  Here we discuss the possibility of a light neutralino decaying via LNV as well.

The LLE signal would look like
Higgs to 4 leptons plus missing energy, where the neutralino decays through a virtual lepton.
For the LQD channel the neutralino may decay either through a virtual slepton or squark.  The signal would be
Higgs to missing energy plus 4 jets, 2 leptons plus four jets, or 4 jets one lepton plus missing energy. All of the signals
for LNV decay of neutralinos may have multiple b squarks.

BNV decays of the Higgs all have a 6 jet topology, however, the flavor antisymmetry of BNV operators does not allow two down type squarks of
the same flavor to be produced by the same vertex.

\begin{figure}[t]
\centerline{\includegraphics{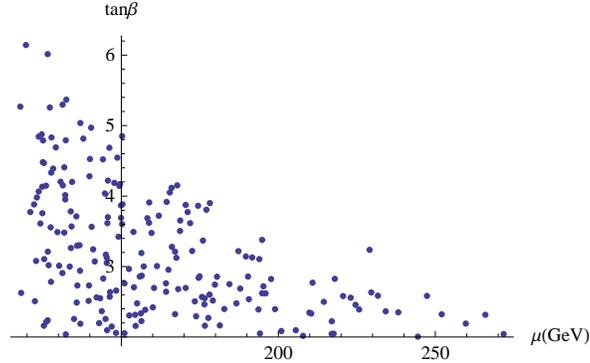}}
\caption{Plot of mu term vs tanb for a Higgs to neutralino cross section which beats Higgs to bbar by a
factor of 5 and satisfies Z width, b to s gamma and chargino mass constraints}
\label{fig:binos}
\end{figure}

\begin{figure}[t]
\centerline{\includegraphics{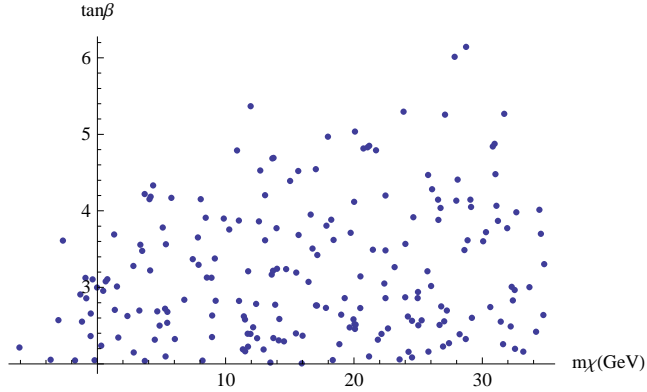}}
\caption{Plot of bino mass vs. $tan\beta$}
\label{fig:binovtanb}
\end{figure}

  Since the decay happens through a virtual heavy squarks or sleptons, we may calculate the decay length.  Here we calculate the decay length for neutralinos decaying via the BNV operator.
\begin{eqnarray}
L & \simeq & \frac{384\pi^2 \cos^2\theta_w}{\alpha \left| U_{21}\right|^2 \lambda^2}\frac{m_{\tilde{m}}^4}{m_\chi^5} ( \beta\gamma)\\\nonumber
& \sim & \frac{3 {\mu}m}{\left| U_{21}\right|^2} \left(\frac{10^{-2}}{\lambda}\right)^2 \left(\frac{m_{\tilde m}}{100\; {\rm GeV}}\right)^4 \left(\frac{30\; {\rm GeV}}{m_\chi}\right)^5 \frac{p_\chi}{m_\chi},
\end{eqnarray}
where $|U_{21}|$ is an element of the mixing matrix  and $p_\chi$ is the neutralino's momentum.  The
neutralino may decay via BNV or either LNV operator.  For large RPV couplings and small offshell masses this decay is prompt.  However if $(\lambda^2/ m_{\chi_0})^4$ is too small the decay will not happen inside the detector and this process will be ruled out by searches for invisible decays of the Higgs.  There is an intermediate range where the decay happens inside the detector but there is a secondary displaced vertex.  In this case, the decay products of each $\chi_0$ will track back to these vertices, which are separate from the  initial vertex where the $\chi_0$'s are made.  To illustrate the point we have constructed a plot assuming the BNV decay of the neutralinos which shows us the RPV coupling vs. offshell squark mass for different displacements of the secondary vertex.  Though it is easy to be in a region of this parameter space with a prompt decay, there is also quite a bit of space that allows for a displaced vertex of a few microns.

\begin{figure}[t]
\centerline{\includegraphics{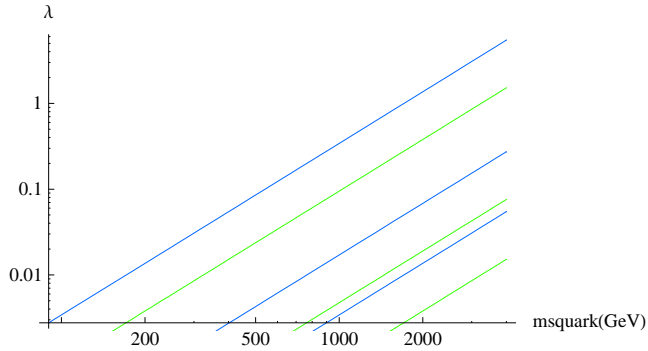}}
\caption{For neutralino masses of 25 and 35 GeV, squark mass vs RPV coupling for displaced vertex of 4 microns, 4cm, and 1m }
\label{fig:squarks}
\end{figure}

Some attention should be paid to the specific issue of like sign dilepton constraints in LNV scenarios.  As mentioned earlier, like sign dilepton measurements force us to consider only decays with taus leptons in the final state.  For decays proceeding through the LQD operator we may easily choose to turn on RPV couplings such that only a tau will appear in the final state.  The LLE operator requires each neutralino decay to two charged leptons and a neutrino.  Since the LLE operator is antisymmetric in left handed fields, and we require the charges leptons to be taus, the neutrino may not be a tau neutrino.  For example the decay may proceed via the coupling 323 to two taus and a muon neutrino.  In this case the coupling 233 has an equal magnitude and the decay products are just as likely to contain a tau a muon and a tau neutrino.  In this case $frac{1}{8}$ of the events will contain like sign (non-tau) dileptons, and if we a standard Higgs production cross section and an efficiency of 8 percent at 1 $fb^{-1}$ we expect to see 10 events over background at 95 percent confidence.  This is within the inclusive excess for the Tevatron search.

We now mention constraints and discuss the parameter space of these decays.

In the case of UDD and LQD operators, there is a
contribution of the Z width to hadrons and the LLE operator contributes to the total Z width. Since our lightest neutralino is mostly Bino, we may sufficiently decouple from the Z to suppress large contributions to the width.  All scans show points within $1\sigma$ of the measured value.  We require that the decay of Higgs to neutralinos beat Higgs $\rightarrow b\overline{b}$ by a factor of 5.  In addition we have used to decoupling of the M1 parameter from M2 to ensure that we may satisfy the chargino lower bound of 102.7 GeV.  We have also plotted points such that the value of $b\rightarrow s \gamma$ does not exceed $2\sigma$ of the measured value while scanning over stop mass parameters and charged Higgs masses.  We have plotted points scanning for a Higgs mass of 91 GeV, however small regions of parameter space exist for Higgs masses as low as 87 GeV.
\subsection{Gluino LSP}

\begin{figure}[t]
\centerline{\includegraphics{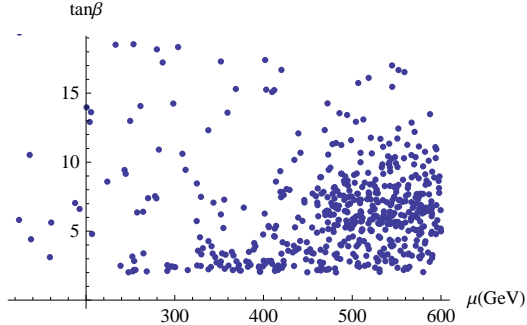}}
\caption{Plot of mu term vs tanb for a Higgs to gluino cross section which beats Higgs to bbar by a
factor of 5 and satisfies Z width, b to s gamma and chargino mass constraints}
\label{fig:gluinos}
\end{figure}

The second possibility for gaugino LSPs are gluinos.  Light gluinos may easily dominate the Higgs decay width as shown in
 \cite{Djouadi:1994js}.  Again, gluinos less than half of the Higgs mass are maximally incompatible with the standard gaugino mass ratio predictions.  However light gluinos do have a model
building benefit, light gluinos do not contribute substantially to squark masses in loops.  When gluinos have mass, there is a two loop gauge mediation-like contribution to squark masses which
 goes like $\frac{\alpha_3}{\pi} M_32 log(\frac{\Lambda}{m})$.  Here $\Lambda$ is the cutoff of the theory, and M3 is the gluino mass parameter.  If the cutoff of the theory is high the squarks get a large mass contribution  and the Higgs sector may become tuned as noted in \cite{Schuster:2005py}.

The Higgs decays to gluinos through a triangle diagram involving top squarks and top quark.  The diagram has a loop factor and is cut off by the largest scale in the loop.  A similar diagram generates Higgs to gluons except only tops or bottoms appear in the loop.  In this case however, the coupling of Higgs to tops is only the Yukawa coupling which is not enhanced by mixing.  The stop's left right eigenstates are not their mass eigenstates.  If the left right eigenstates have a large mixing, one stop will be much lighter than the other.  In addition the Higgs coupling to $t_{1}t_{1}$  is proportional to the off-diagonal element of the mass matrix $\sim m_{t1}^2-m_{t2}^2$.  As the coupling becomes large, this decay may easily beat Higgs $\rightarrow b\overline{b}$. See figure \ref{fig:gluinos} for the stop sector parameters needed to beat the bbar cross section by a factor of 5, where we have plotted parameter space in the limit of light gluinos $\sim 20 GeV$ or less and a 100 GeV Higgs.

The gluinos are then free to decay via an LNV or BNV operator. The topology of this decay is similar to that of the neutralino, the decay will proceed trough a virtual squark, and result in six fermions in the final state.  In the case of BNV, the gluino will decay to a quark
and an offshell squark, which will then decay to two squarks; thus the signal is Higgs to six jets.  If the gluino were to decay via an LNV vertex, it would be via the LQD operator.
 In this case the gluino would decay to a quark and an offshell squark.  The squark then decays to a quark and lepton.  The signal would then be Higgs to 4 jets plus missing energy or Higgs to 4 jets plus 2 charged leptons.  Because there is no flavor antisymmetry in the LNV operator, these signals may contain multiple quarks of heavy flavor.  In particular, if the final state leptons are neutrinos, we may have the signals 4b plus missing energy, 4c plus missing energy or 2b and 2 c plus missing energy(since each neutralino is free to decay to bottom or charm pairs).   Perhaps the most striking signal would contain 2b, 2c and 2$\tau$.  There may also be signals in which one gluino decays with a neutrino in the final state, while one decays with a tau.  In this case the decay products may contain an odd number of a heavy quark such as 3b, c, tau,and missing energy or 3c, b, tau and missing energy.

  One might worry that a similar process also contributes to Z to gluinos which would negatively effect the measured $e^+ e^- \rightarrow$ hadrons, and the total hadronic width of the Z.  However the coupling of stops to the Z is only a gauge coupling which can't compete with the coupling to the Higgs.  Further, the coupling of the stops or sbottoms (but not both at the same time) to the Z may be tuned away while remaining large for the Higgs.   The $b \rightarrow s\gamma$ measurement mostly constrains allowed values of $tan\beta$, as part of the $b \rightarrow s\gamma$ amplitude goes like $1/cos\beta$.  Our plot shows parameter space for which the $b\rightarrow s \gamma$ prediction is within 2$\sigma$ of the measured value.
  Like the neutralino, since the decay of the LSP proceeds through an offshell sparticle, there exists the possibility of a secondary displaced vertex.  For sufficiently large RPV coupling and small mass of the offshell sparticle, the decay will occur inside the detector.  In addition, we note that the decay length in this scenario will be more likely to be shorter than that of neutralino LSPs, as it goes like $\frac{1}{\alpha}$.  Finally, as is the case with neutralinos, like sign dilepton constraints require that the LNV operator must have a structure that favors third generation couplings.

\section {Scalars}
\subsection{Topology and General Constraints}
The Higgs may decay to a pair of scalars, which then each decay directly through an R parity violating operator.  These decays of the Higgs are usually prompt, and have a 4 particle final state topology.
These scenarios are constrained by the chargino lower mass bound and $b\rightarrow s\gamma$ as were the gaugino LSP scenario.  Since the scalar LSP pair must have opposite charges, like sign dileptons do not constrain this scenario.  Again we must insure that these decays can beat Higgs $\rightarrow b\overline{b}$ by a suitable factor.  However, as we explain, mass bounds on the scalar LSPs themselves are of the biggest concern.

Several experiments have put lower mass bounds or production cross section limits on scalars decaying through RPV.  The least stringent bounds for $\tilde{\tau}$ decay are from L3 and OPAL respectively and are $m_{\tilde{\tau}}> 61$ for LLE and $m_{\tilde{\tau}}> 74$ for LQD couplings.   For sbottoms the least stringent limits quoted by L3 for UDD decays and is $m_{\tilde{b}}> 55$.  These limits would seem to exclude sbottom and stau as LSPs in our scenario.  However none of these searches looked for RPV decays of these particles in mass windows much below half of the Z mass. Table 1 compiles a list of the lowest masses used in direct scalar searches  \cite{Abdallah:2003xc},\cite{Achard:2001ek}(table 3),\cite{Abbiendi:2003rn},\cite{Heister:2002jc}.  It was assumed that particles less than half of the Z mass could be ruled out with Z width measurements and with contributions to $e^{+}e^{-}\rightarrow hadrons$.

As we shall elaborate, third generation scalars may have substantial mixing in their mass matrices.  By varying the scalar mixing angle, we may change their coupling to the Z.  By choosing a small coupling to the Z, we may suppress contributions to $e^{+}e^{-}\rightarrow hadrons$ and the Z width for scalars with masses below half of the Higgs mass.  In this way we may avoid RPV lower mass bounds for third generation scalars.

For scalars, the mass eigenstates and left right eigenstates are not the same.  We may write,
\be
s1= Sin\theta s_L + Cos\theta s_R
\ee
\be
s2= Cos\theta s_L-  Sin\theta s_R
\ee
where $\theta$ is the $s_1$ $s_2$ mixing angle.  The mass matrix for third generation scalars is given by,

\centerline{$\left(
\begin{array}{cc}
M_{sL}^2 + ms^2 + D_{\tilde s_L} & m_s A \\
 m_s A & M_{sR}^2 + ms^2 + D_{\tilde s_R}\\
\end{array}
\right)$}

where $A=A_{su}-\mu cot\beta$ for up-type, $A=A_{sb}-\mu tan\beta$ for down-type.  The mixing angle is

\be
sin2\theta =2 m_b A/(m_{s1}^2-m_{s2}^2)
\ee

For third generation sparticles it is possible to get large off-diagonal terms and we may have one mass eigenstate which is  much lighter than the other.  In general, it is not too hard to achieve a light mass eigenstate less than half of the Higgs mass.  In addition the couplings of the light mass eigenstate to the Z are controlled by the mixing angle, \be
g_{Zs_2 s_2}=g(I_3 sin_w^2-Q_2 sin^2\theta)
\ee
which may be adjusted to vanish.  Due to differing charges, the angle at which the scalar completely decouples will be different for stops, sbottoms and staus.  It is worth noting that though third generation scalars can be decoupled from the Z, nothing will decouple them from photons. Thus while we can suppress the contribution of scalars to $e^+ e^- \rightarrow hadrons$, we cannot eliminate it.  Henceforward we will concentrate on Higgs decaying through stau and sbottom LSPs.

\begin{table}
\label{tab:points}
\begin{center}
\begin{tabular}{c|c|c|c}
EXP &  LLE & LQD & UDD\\
\hline

DELPHI & $\tau >$ 45 & - & b $>$ 45\\
OPAL & $\tau > $ 45 & $\tau >$ 45   & b $>$ 45 \\
L3 & $\tau >$ 70 & - & b $>$ 30 \\
ALEPH & $\tau >$ 45 & $\tau >$40, b$>$ 30 & b $>$ 45 \\
\hline
\hline
\end{tabular}
\caption{LEP2 experiment searches for scalars decaying directly through RPV}
\end{center}
\end{table}

\subsection{Sbottom LSP}

The bottom squark is a good candidate for a light LSP.  Its decay may proceed either through LNV or BNV operator.
In the case of BNV the decay proceeds as Higgs to light sbottoms with each light sbottom decaying to an up
 and down quark directly through the RPV coupling.  Because of the down type flavor antisymmetry no bottoms appear
in the final state.  The signal is thus Higgs to 4 jets, at most 2 of heavy flavor(charm).  LNV decays proceed through the LQD operator, with the signal two quarks and two leptons.  In the case of the LQD operator there is no flavor antisymmetry, thus there may be a b and a $\overline{b}$ in the final state.  If the final state leptons are neutrinos, the most visible decay will be Higgs to 2b plus missing energy.   The most visible signal in this case would be $h \rightarrow 2c + 2\tau$.

\begin{figure}[t]
\centerline{\includegraphics{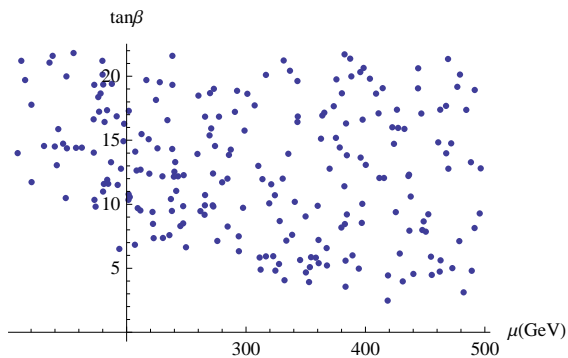}}
\caption{Plot of mu term vs tanb for a Higgs to sbottom cross section for sbottoms between 7.5 and 30 GeV which beats Higgs to bbar by a
factor of 5 and satisfies Z width, b to s gamma and chargino mass constraints}
\label{fig:sbs}
\end{figure}

We will now discuss constraints and the parameter space.  Long lived sbottoms are ruled
out under 92 GeV.  However, as long as the RPV couplings are large enough and the sbottom is sufficiently more massive than its decay products, the decay should be prompt.
We see that by adjusting the mixing angle of the sbottoms, we decrease the coupling to Z.  Following the formula quoted in the beginning of this section, we see that this coupling is turned off when $sin\theta \sim .39$.  A lower bound on sbottom masses
has been set at 7.5GeV by measuring contributions to the over
all cross sections of $e^+e^- \rightarrow hadrons$ while turning off the sbottom coupling to Z's \cite{Janot:2004cy}.   As quoted in table 1, the L3 experiment put an effective upper bound of 30 GeV on sbottom squarks decaying through RPV.  Previous bounds had been set on the mass splitting of very light sbottoms and the lightest stop \cite{Carena:2000ka}.  These took into account a large stop loop contribution to the Higgs mass and limits on the $\rho$ parameter and required a light stop lighter than 300 GeV.  In our scenario however the bound is even more relaxed since we do not require a large stop loop contribution to the Higgs, and our stop sbottom splitting is not quite as extreme.

The decay rate of Higgs to sbottoms must beat that of bottoms by a factor of a few.  The ratio of decays rates is given by \cite{Berger:2002vs} as
\be
 \Gamma_{\tilde{b}}/ \Gamma_b= \mu tan\beta^2 / 2 m_h^2 sin2\theta^2 (1-4m_{b}^2/m_h^2)^{1/2}
\ee
 which gets large for large values of $\mu$ and tan$\beta$.  In this scenario there are 5 free parameters; the A term, $\mu$, tan$\beta$, and the soft masses.  We have plotted $\mu$ vs tan$\beta$ in parameter space for this window of allowed sbottom mixing angles.  We see in the plot of that lower values of $\mu$ and tan$\beta$ are ruled out by the upper mass limit on the sbottom.  If the product $\mu$tan$\beta$ is too small, the off diagonal elements of the mass matrix will not be big enough to produce a light mass eigenstate.  We therefore expect to find parameter space at larger values of $tan\beta$ than we did for neutralino LSPs.  The measurements of $e^{+}e^{-} \rightarrow hadrons$ constrain allowed values of $\theta$.  Janot tells us the largest allowed mass range for sbottoms occurs between mixing angles $.3 < sin\theta < .45$ \cite{Janot:2004cy}(see fig 8).  For A terms and $\mu$ terms of a few hundred GeV we see that we may fall into this range of mixing parameter if A and $\mu$ cancel to within 20 or 30 percent of their value.  The constraint from b to s gamma follows as before and we scan over stop sector parameters and obey the chargino constraint. However neither of these constraints is as restrictive as for the case of light neutralinos since we are free to make the gaugino sector heavier in general.

\subsection{Stau LSP}

The decay of the Higgs may proceed through a stau anti-stau LSP pair.  The direct RPV decays of the tau happen through LNV operators only.

For LQD decays of the stau, the final state will have 4 quarks, 2 up type and 2 down type.  Because there is no flavor antisymmetry, the final state may consist of all quarks of heavy flavor, $h \rightarrow b\overline{b}c\overline{c}$.

\begin{figure}[h]
\centerline{\includegraphics{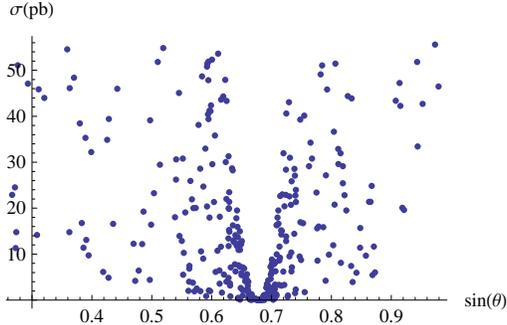}}
\caption{Plot of stau mixing vs hadronic cross section at the Z pole.  The lower region is ruled out by PEP PETRA and TRISTAN measurements, the regions to the left and right by Z pole data from LEP 1.}
\label{fig:hadrons}
\end{figure}

\begin{figure}[h]
\centerline{\includegraphics{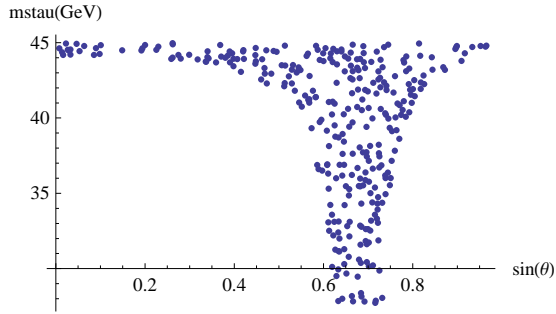}}
\caption{Plot of allowed stau mass vs. stau mixing angle}
\label{fig:massangle}
\end{figure}

 For decays with quarks in the final state, the $\tilde{\tau}$ masses below half the Z mass are constrained by the measurement of $e^{+} e^{-} \rightarrow hadrons $.  Following the method of Janot for light sbottoms, we may go back to LEP2, PEP, PETRA, and TRISTAN measurements of the total $e^{+} e^{-} \rightarrow hadrons$ cross sections to set lower mass bounds on the staus \cite{Janot:2004cy}.  Ref \cite{Janot:2004cy} contains a compilation of the total $\sigma_{had}$ measurements for various experimental energies.  We first calculate the absolute minimum allowed $\tilde{\tau}$ mass by assuming complete decoupling from the Z and calculating the resulting contribution to $e^{+}e^{-} \rightarrow hadrons$ for production from photons alone for different experimental energies.  We then perform a $\chi^2$ fit for the hadronic cross section contribution.  We find that at 95 percent confidence the lower limit on stau masses is 11 GeV with a best fit value of 57 GeV.  A plot of the LEP and low energy cross section measurements and cross section predictions for light staus appears in the appendix.  There seems to be significant parameter space for light stuas.  We must note that Janot performed a more sophisticated analysis of the low energy data which may further constrain the lighter stau masses.  However for stau mass heavier than 28 GeV the lowest energy data does not constrain our scenario.  We will therefore conservatively consider the viable allowed stau mass window to be above 28 GeV.

In addition we may analyze the Z pole data from LEP 1.  As quoted by Janot, the Z pole measurements limit new contributions to the hadronic cross section to 56pb at 95 percent confidence.  Taking these measurements into account, we may exclude stau masses for different couplings of staus to the Z.  When the $\tilde{\tau}$ has small coupling to the Z, we find a mass window  between 28 and 45 GeV.  Following the formula in section 4, we see that the Z coupling to $\tilde{\tau}$ is turned off when $sin \theta=.67$, and allowed mixing angles range from  $.6< sin\theta < .7$ in the case of light staus.  We have plotted the total contribution from light stau production at the Z pole vs mixing angle, as well as allowed stau mixing angles vs stau mass taking into account both Z pole and PEP,PETRA and TRISTAN data.

\begin{figure}[h]
\centerline{\includegraphics{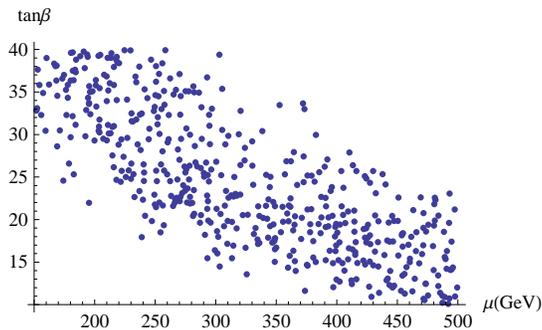}}
\caption{Plot of mu term vs tanb for a Higgs to staus cross section, with a Higgs mass of 105 GeV, which beats Higgs to bbar by a
factor of 5 and satisfies Z width, b to s gamma and chargino mass, and hadronic cross section constraints}
\label{fig:stauspace}
\end{figure}

 In principle the parameter space for this decay looks much like the parameter space for sbottoms.  However, the constraint on the allowed mixing angle is tighter since a large contribution of $e^+e^- \rightarrow hadron$ comes directly from the photon coupling to tau with no coupling to the Z at all.  Again getting a light stau requires large contributions to the off-diagonal mass matrix elements, in this case given by $m_{\tau}(A-\mu tan\beta)$.  Since $m_{\tau}$ is small, a large product $\mu$ $tan\beta$ is required for large mixing, even larger than what was required for light sbottoms.  We keep in mind that we do not want excessively high values of $\mu$, since this would again introduce tuning into the Higgs potential.  In addition tan$\beta $ must not become too large or the bottom Yukawa coupling will become nonperturbative.  We also need to choose $\mu$ and $tan\beta$ such that we may obey the lower mass bound on the lightest chargino.  In this case, the variables effecting the stau sector that are at play in the $b \rightarrow s\gamma$ process are only $tan\beta$ and $\mu$, the stau and stop sectors are fairly uncoupled and $b \rightarrow s\gamma$ is not a strong constraint in the light stau scenario.  We have plotted parameter space for all constraints.

The LLE decay of the stau is an interesting subject.  If allowed, each stau would decay into a charged lepton plus a neutrino, and the signal would be opposite sign dileptons plus missing energy.  The most striking signal here would be $h \rightarrow 2\tau + \not{E}$.  A search with similar signature, gauge mediated decays of stau to tau plus gravitino \cite{Abreu:2000nm}, highly constrains this scenario.  This search rules out the tau plus missing energy signal for stau mass larger than 2GeV.  It should be noted that this search found a small window for $m_{\tau} < m_{\tilde{\tau}} < 2 GeV$.  One might guess that in the case where both staus decay hadronically, contributions to $e^{+}e^{-} \rightarrow hadrons$ rule out this possibility.  However, we  see that the PEP, PETRA, and TRISTAN data placed cuts specifically to rule exclude the background from $\tau^{+}\tau^{-}$ and it is likely these decays may have been missed.  See for example \cite{Von:1990}).  Thus there may be a signal for Higgs to 2 stau plus missing energy in the improbable event that there exist super-light staus.  In this case, since there is a only a small window between the tau and the stau, if the RPV coupling is small, the stau may live for some time before it decays.  In this case, in addition to a 2$\tau$ plus missing energy decay, they may also be a secondary displaced vertex.  For example for LLE coupling of the size a few times $10^{-4}$ a 1.8 GeV stau may live for 100 microns.  Stau pair decays with one or two light leptons may also occur.

\section{Conclusions}
In the R parity violating MSSM without standard gaugino mass relationships, we find a menagerie of new decays for a light Higgs.  We present the list of possible LSPs and topologies in the table found above.

One may wonder how all of these signals pass current bounds.  There are several current searches that one must consider and we will first mention general of these.  DELPHI did an analysis for Higgs goes to anything which could put a lower mass bound on the Higgs of 82 GeV \cite{Abbiendi:2002qp}.  There has been a Higgs to missing energy search which put a lower bound on the Higgs mass of 114 GeV \cite{Schael:2006cr}.  This is a concern
for us only when the Higgs decay is not prompt as is the case for gaugino LSPs.  It is easy to avoid falling into this search if RPV couplings are large or if squark or slepton masses are of reasonable size.  In fact, in the case of squarks, we would not want masses to go much beyond 1TeV, for then we encounter the same tuning problem in the Higgs potential that we wished to avoid.

We shall now consider searches which apply to decays with all hadronic final states.  First there is the 2 jet flavorless search which puts a lower bound on the Higgs of 113 GeV \cite{:2001yb}.  In order for this search to be sensitive to our scenario, we would have to force our 4 and 6 jet final states into two jets.  Such forcing is used in a set of searches where members of the Higgs multiplet undergoes cascade decay \cite{Abbiendi:2004ww} \cite{Abdallah:2004wy}.  Thus we worry about decays $e^+e^- \rightarrow H_2Z \rightarrow H_1H_1Z$, where $H_1$ decays to bottoms or taus.  In the all hadronic case, the final state is 4b.  The event is forced into two 2b jets and the efficiency is calculated of a 4b even being picked up by a 2b search.  Using the DELPHI search table we can guess our efficiency of being picked up by the 2b search.  This should give us a good idea about matching our 4 jet signals to 2 jets, for 6 jets the efficiency would be even worse.  As an example, notice that to rule out the Higgs at 80 GeV, the 4b search needs about 5.5 times the number of events as the 2b.  Even assuming both of our b's were tagged, our scenario would only be picked up 18 percent of the time.  The flavorless search is the same as the 2b search without the b tags\cite{:2001yb}.  LEP performed a search in which HZ was produced and Higgs decayed to $WW^{*}$ \cite{Schael:2006ra}.  An analysis was done for the final state $Z \rightarrow \nu\nu, H \rightarrow q\overline{q}q\overline{q} $, which is relevant to our light stau scenario.  This search places a lower mass bound on the Higgs of 105 GeV.  In addition the $WW^{*} $ search considered a final state in which $e^+e^- \rightarrow  HZ\rightarrow 6q$.  In this case however, cuts are applied which reconstruct the masses of the final state Z and Ws, so this search is not likely to be sensitive to our 4 or 6 jet signals.

We now turn to searches with quarks and charged leptons in the final states.  The cascade decays mentioned above fall into this category produce final states with 4$\tau$ and 2b+2$\tau$.  We can compare these final states to all of ours and see that these searches are not directly sensitive to any of our final states.  Of our final states without missing energy these signals come closest are the six body decay 2b+2c+2$\tau$ and the four body decay, 2c+2$\tau$.  Reconstruction of the cascade decay requires multiple b tags.  It is unlikely that the 2 charms will receive a large b-likeness parameter.  In addition to the b tag problem, our 2b+2c+2$\tau$ decay must be force four jets into two loosing efficiency.  We do not expect that these searches will constrain our scenario.  Cascade decays with even more final state b's and $\tau$s occur through the process $e^+e^- \rightarrow  H_2H_2\rightarrow H_1H_1H_1$.  Again, none of these final states match directly to our signals.  Events with 4 or 6 b's require even more b tags, which are unlikely to match up to our 6 particle final states that have at most 2 b's and none of our final states contain more than two taus.

Finally there are decays with missing energy in the final state.  A compendium of some searches that constrain Higgs signals with missing energy is found in \cite{Chang:2007de}.  In particular this paper quotes upper bounds on the Higgs production cross section for Higgs decays with final states and $2q + \not E$.  These limits do not come from a Higgs search but from the LEP2 squark searches where sparticles are pair produced and decay to quarks and neutralinos.  This signal is identical to our scenario with quarks and missing energy in the final state.  In this case a lower bound may be set on the Higgs mass of 103 GeV if the final state quarks are light and 111 GeV if both of the final state quarks are b quarks.  These bounds do not apply in cases where the final state quarks are not the same flavor.  One might consider the cascade decays $e^+e^- \rightarrow H_2Z \rightarrow H_1H_1Z\rightarrow 4b/4\tau$ $Z$.  If in this case the Z decays invisibly the overall signal would be identical to our $H \rightarrow 4b/4\tau + 2\nu$ final state.  The 4b search involved cuts which considered parameters such as the missing mass and the $ln(\chi_{mZ})$ parameter where the missing mass is forced to the Z mass.  These cuts are bound to exclude data in our scenario.  Further we can see from the exclusion plot in fig 12 that if the efficiency of observing out scenario is 60 percent, all of parameter space is open up down to the 82 GeV Higgs mass bound.  The $4\tau$ search constraints on the decay topology such as restrictions on the angle between charged particle pairs, in addition to cuts on the missing momentum which make it insensitive to our decay.  However the in the $WW^{*}$ search an analysis was done for the channel $e^{+}e^{-} \rightarrow HZ \rightarrow WW^{*}Z$ where the Ws decay hadronically and the Z invisibly.  Here we cannot avoid constraints and a lower limit of 105 GeV may be placed on the Higgs mass.  Finally, the $WW^{*}$ search analyzed a Higgs channel with charged light leptons and missing energy in the final state.  This signal is identical to out final state where scalar LSPs decay through the LLE operator.  Even in the case that we consider an LLE operator like 313, where the final state leptons may be only taus, the antisymmetry of the RPV operator forces us the coupling 133 to be the same size. Thus we get light leptons a quarter of the time and this search will also constrain the 2 tau plus missing energy channel.  The search places an upper bound on the Higgs production cross section of .044pb and in our case, this translates to a Higgs mass lower bound of around 104 GeVfor final states with taus or light leptons plus missing energy.  We may lower the bound by turning on more than one RPV coupling at once and decreasing the likelihood that the decay products are symmetric.  For example, turning on two coupling at once which are equal in size decreases the Higgs mass bound to 95 GeV.

Other asymmetric decays are also listed in table 2 which involve 2 RPV coupling being turned on at once.  These scenarios are bizarre enough to be mostly unconstrained.  The final state b+c+$\tau +\nu$ is mentioned in the $WW^{*}$ search, however the combinatorics of turning on two RPV operators at once allow us to avoid the Higgs mass bound of 95 GeV set by this scenario.  Stranger 6 body decays listed in our table have are not constrained by any relevant search.

\begin{table}
\label{tab:points}
\begin{center}
\begin{tabular}{c|c|c|c}
 LSP &  LLE & LQD & UDD\\
\hline

$\chi_0$ & 4$\tau$+$2\nu$  & 4b/4c+2$\nu$, 2b+2c+2$\nu$, 2b+2c+2$\tau$,3b+c+$\tau$+$\nu$, b+3c+$\tau$+$\nu$ & 2b+2c+2q\\
g & - & 4b/4c+2$\nu$, 2b+2c+2$\nu$, 2b+2c+2$\tau$,3b+c+$\tau$+$\nu$, b+3c+$\tau$+$\nu$   & 2b+2c+2q \\
b & - & 2b+2$\nu$, 2c+2$\tau$, b+c+$\nu$+$\tau$ & 2c+2q  \\
$\tau$& 2$\tau$+2$\nu$ & 2b+2c & - \\
\hline
\hline
\end{tabular}
\caption{Higgs decay signals for all possible LSPs and RPV operators}
\end{center}
\end{table}

\begin{table}
\label{tab:points}
\begin{center}
\begin{tabular}{c|c|c|c}
 LSP &  Signature & Mass Bound & Search\\
\hline

$\chi_0$ & $4q+2\nu$ & 105 GeV & $WW^{*}$ with invisible Z decay\\
$\tilde{g}$ & $4q+2\nu$ & 105 GeV &  $WW^{*}$ with invisible Z decay \\
$\tilde{b}$ & $2q+2\nu$ & 103 GeV & SUSY squark search\\
- & $2b+2\nu$ & 111 GeV & SUSY squark search\\
- & $4q$ & 105 GeV & $WW^{*}$ with invisible Z decay\\
$\tilde{\tau}$ & $\tau\overline{\tau}+2\nu$ & 104 GeV & $WW^{*}$ \\
- & $l\overline{l}+2\nu$ & 104 GeV & $WW^{*}$ \\
- & 4q & 105 GeV &$WW^{*}$ with invisible Z decay \\

\hline
\hline
\end{tabular}
\caption{Higgs Mass Lower Bounds for Various Channels. For decays not listed current searches do not severely constrain the Higgs mass.}
\end{center}
\end{table}

Overall we have proposed over a dozen distinct channels for Higgs discovery with multiple heavy flavor particles in their final states.  Some of these decays would be very hard to detect, for example those decays in which the final state is 4 or 6 jets.  However, many of these decays have interesting missing energy signatures, some of which are quite bizarre -  for example the b+3c+$\tau$ +$\nu$ decay of the gauginos.  Those decays that proceed through a gaugino LSP have the added bonus of possible secondary displaced vertices.  We may imagine modifying existing searches to look in some of these channels, for example modifying existing 2 or 4 b and $\tau$ searches to be sensitive to missing energy.  In addition we might hope to detect events which contain bottoms at LHCb as has been recently proposed \cite{Kaplan:2007ap}.

\section{Appendix}

Below we have plotted hadronic cross section vs center of mass energies.  The points are the measurement of the hadronic crossection minus the SM theory prediction with two sigma error bars.  The curves are contributions to the hadronic cross section from the decay of low energy staus of various masses.

\begin{figure}[h]
\centerline{\includegraphics{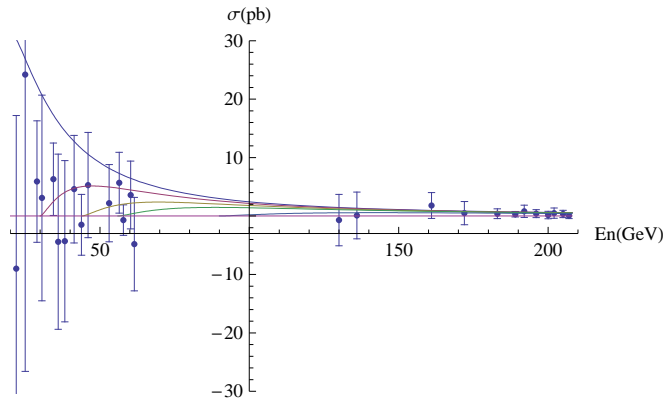}}
\caption{Plot of Energy vs hadronic cross section for low energy data and stau contributions.  from the top down the curves are for staues of mass 10, 15, 22, 28, and 45 GeV }
\label{fig:stlength}
\end{figure}

{\bf Acknowledgments}

This work was supported in part by funding from the US Department of Energy.  We would like to thank Tom Banks and Jason Neilsen for enlightening discussion about signals and constraints.

\bibliographystyle{apsrev}

\end{document}